\documentclass[final,journal]{IEEEtran}

\usepackage{graphicx}  
\usepackage{amssymb,amsmath}
\usepackage{tikz}\usetikzlibrary{arrows,shapes,trees}
\usepackage{cite}
\usepackage{bm}

\title{Feed-Forward Staircase Codes}
\author{Lei M. Zhang and Laurent Schmalen
\thanks{The associate editor coordinating the review of this letter and approving it for    publication was Dr. xxx. Manuscript received XXX. yy, 2016.}
\thanks{L. M. Zhang is with the University of Toronto, ECE department. His work has been carried out while he was visiting Nokia Bell Labs funded by a scholarship from the German DAAD-RisePro program.}
 \thanks{L. Schmalen is with Nokia Bell Labs, Stuttgart, Germany (e-mail: first.last@nokia.com).}
\thanks{L. Schmalen was supported by the German BMBF in the scope of the CELTIC+ project SENDATE-TANDEM.}
\thanks{Digital Object Identifier xx.xxxx/xxx.2016.xxxxxx}
}

\newcommand{\bI}{\bm{I}}
\newcommand{\bF}{\bm{F}}
\newcommand{\bG}{\bm{G}}
\newcommand{\bA}{\bm{A}}
\newcommand{\bB}{\bm{B}}
\newcommand{\bC}{\bm{C}}
\newcommand{\bE}{\bm{E}}
\newcommand{\bM}{\bm{M}}
\newcommand{\bP}{\bm{P}}
\newcommand{\bX}{\bm{X}}
\newcommand{\bY}{\bm{Y}}
\newcommand{\bQ}{\bm{Q}}
\newcommand{\bPi}{\bm{\Pi}}

\begin{document}

\maketitle
\begin{abstract}
We propose two variants of staircase codes that resolve the issue of parity-propagation in their encoding process. The proposed codes provide a systematic way of terminating a staircase code after an arbitrary number of blocks. The class of feed-forward staircase codes are introduced, which uses a self-protection technique to avoid parity-propagation. We also introduce the class of partial feed-forward staircase codes, which allows parity-propagation to occur over a given number of blocks. By amortizing the complexity of self-protection over several standard staircase blocks, the encoding complexity of these codes is made comparable to staircase codes. Partial feed-forward staircase codes have the same error-floor as staircase codes. Simulations confirm that the performance of the proposed codes in both the waterfall and error-floor regions is similar to the original staircase codes. The proposed codes help extend the domain of application of staircase codes to systems in which parity-propagation is undesirable or termination is necessary.
\end{abstract}

\section{Introduction}

\IEEEPARstart{H}{igh-speed} fiber optical communication system are a challenging environment for {forward error correction} (FEC) schemes. Modern high-speed optical communication systems require high-performing FEC engines that support throughputs of 100 Gbit/s and multiples thereof, that have low power consumption, that realize  net coding gains (NCGs) close to the theoretical capacity limits at a target BER of $10^{-15}$, and that are preferably adapted to the peculiarities of the optical channel~\cite{Leven2014status}. 

Although coding schemes that allow for soft-decision decoding are now well established in optical communications~\cite{Leven2014status}, especially in long-haul and submarine transmission systems which need to operate at the lowest possible {signal-to-noise ratio} (SNR), hard-decision decoding is still predominant in the widely deployed metro networks, due to its low complexity leading to power-friendly receiver implementations~\cite{pillai2014end}. Such low-complexity receivers are also attractive for data center interconnect applications.

In the recent years, several new capacity-approaching coding schemes suitable for high-speed optical communications have been presented. Staircase codes \cite{smith:2012,zhang:2014}, are hard-decision decoded, spatially-coupled codes with practical application in forward error-correction for long-haul optical-fiber transmission systems. An ITU-T G.709-compatible staircase code with rate $R=239/255$ was shown to operate within 0.56\,dB of the capacity of the binary-input AWGN channel with hard decision at the output (which is equivalent to a binary symmetric channel (BSC)) at a bit-error rate (BER) of $10^{-15}$ \cite{smith:2012}. Its gap to capacity is smaller than all of the enhanced coding schemes proposed in ITU-T recommendation G.975.1 \cite{g.975.1:2004}. In \cite{zhang:2014}, staircase codes with rates $R \geq 6/7$ were shown to be within 0.80 dB of capacity in terms of NCG at a BER of $10^{-15}$. Such coding gains are obtained by using an iterative, hard-decision decoding algorithm with decoder data-flow orders of magnitude lower than that of message-passing decoders for sparse-graph codes such as Turbo or Low-Density Parity-Check (LDPC) codes \cite{smith:2012}. For long-haul optical-fiber transmissions systems where bit-rates exceed 100 Gb/s, staircase codes are often the best practical solution.

Besides staircase code and variants thereof~\cite{hager_2015}, several other code constructions based on spatial coupling of algebraic component codes have been proposed, e.g., braided BCH codes~\cite{jian:2013}. Recently, multiple works show that these codes can approach capacity of the BSC under simple iterative decoding when the rate is large enough~\cite{jian:2012,jianthesis:2013,hager2015density,zhang2015spatially}. However, all the proposed structures of spatially coupled algebraic product codes are recursive codes which lead to several practical drawbacks in their implementation: First, a recursive structure requires extra circuitry~\cite{TavaresPhD} for terminating the code, which may be undesired in some applications where a low-complexity decoder implementation is crucial. Previous publications have not explicitly dealt with code termination but have only considered free-running, non-terminated codes. Terminating a feed-forward code on the other hand is straightforward. 

A second drawback of recursive codes is the effect of \emph{parity-propagation}; a single non-zero information bit leads to an infinitely extending parity sequence. This effect may be undesired in some optical transmission applications, where the transceivers are usually free-running due to the setup times of links~\cite{wu2011survey} and only some of the transmitted bits carry useful information. Parity propagation limits in this case the possibility of switching off the forward error correction circuitry during times when no useful data is transmitted, non-negligibly increasing the transceiver power consumption~\cite{pillai2014end}.

In this paper, we discuss several options for constructing feed-forward staircase codes. It becomes quickly obvious that a straightforward modification of the staircase encoding structure to avoid parity propagation will lead to unacceptably high error floors for most applications. In order to mitigate the error floor, we use the technique of self-protecting parity-bits~\cite[App. I.9]{g.975.1:2004} together with a clever interleaving to construct a class of feed-forward staircase codes. We also give an approximation of the expected error floor based on the minimum size stall pattern. In some applications with very stringent requirements, the error floor may still be too large. For this reason, in the second part of the paper, we slightly relax the parity-propagation constraint and present \emph{partial feed-forward} staircase codes, which have the same error floor as the original staircase codes but completely avoid parity-propagation and allow for easy termination.

This paper is structured as follows: In Sec.~\ref{sec:background}, we introduce the basic notation and recapitulate the structure and main properties of staircase codes. In Sec.~\ref{sec:ff_sc}, we introduce a first construction of feed-forward staircase codes based on self-protected parity-bits. In Sec.~\ref{sec:pff_sc}, we slightly generalize this construction and introduce partial feed-forward staircase codes, which have a slightly lower rate but improved error floor properties. Error floor approximations based on minimal stall patterns are derived in Sec.~\ref{subsec:error_floor}. Finally, we compare in Sec.~\ref{sec:simulations} the performance of both schemes using a coding setup typically used in optical communications.

\section{Background: Staircase Codes}\label{sec:background}

In this section, we briefly overview the encoding and decoding of staircase codes since the proposed codes share many common features with the original staircase code. We also motivate our work by examining the parity-propagation property of staircase codes. 

\subsection{Notation}

Given integers $a$, $b$ where $a<b$, let $[a,b] \triangleq \{a,a+1,\dots,b\}$. For an $m \times n$ matrix $\bm{Q}$, we denote a vectorization of $\bm{Q}$ by $\textrm{vec}(\bm{Q})$, where the resulting vector is assumed to be a column vector and the mapping between matrix and vector indices is given by a bijection $v : [0,m-1]\times[0,n-1] \to [0,mn-1]$. The inverse of $\textrm{vec}(\cdot)$ is denoted by $\textrm{vec}^{-1}(\cdot)$ with the underlying mapping $v^{-1}$, the inverse of $v$. For example, the mappings of the column-wise vectorization and its inverse are given by
\[
v(i,j) = jm+i\quad\text{and}\quad v^{-1}(i) = (i \textrm{ mod } m,\lfloor i/m \rfloor).
\]
We denote the $m \times 1$ unit vector with a single $1$ in the $i$th position by $\bm{e}_{m,i}$. We denote the $m \times m$ identity matrix by $\bI_{m}$ and the $m \times n$ all-zeros matrix by $\bm{0}_{m \times n}$. Let $\bE_m$ denote the $m \times m$ elementary permutation matrix, obtained by cyclically shifting each row of $\bI_m$ to the right by $1$. Recall that for $i \geq 0$, $\bE_m^i$ is a permutation matrix obtained by cyclically shifting each row of $\bI_m$ to the right by $i$.

Given an $m \times n$ matrix $\bA$ and another matrix $\bB$, their Kronecker product is defined as
\[
\bA \otimes \bB \triangleq \left[
\begin{array}{ccc} 
a_{00}\bB 	& \dots 	& a_{0(n-1)}\bB \\
\vdots 		& \ddots & \vdots \\
a_{(m-1)0}\bB & \dots 	& a_{(m-1)(n-1)}\bB
\end{array}
\right].
\]
A block diagonal matrix consisting of $m$ copies of a matrix $\bm{Q}$ along its diagonal is given by $\bI_m \otimes \bm{Q}$. We denote a block diagonal matrix consisting of $m$ arbitrary matrices $\{\bm{Q}_1,\bm{Q}_2,\dots,\bm{Q}_m\}$ of the same size along its diagonal by
\[
\bm{\mathcal{B}}(\bm{Q}_1,\bm{Q}_2,\dots,\bm{Q}_m) \triangleq \sum_{i=1}^m (\bm{e}_{m,i}\bm{e}_{m,i}^T) \otimes \bm{Q}_i.
\]

\subsection{Encoding of staircase codes}\label{subsec:staircase_enc}

\begin{figure}
\includegraphics{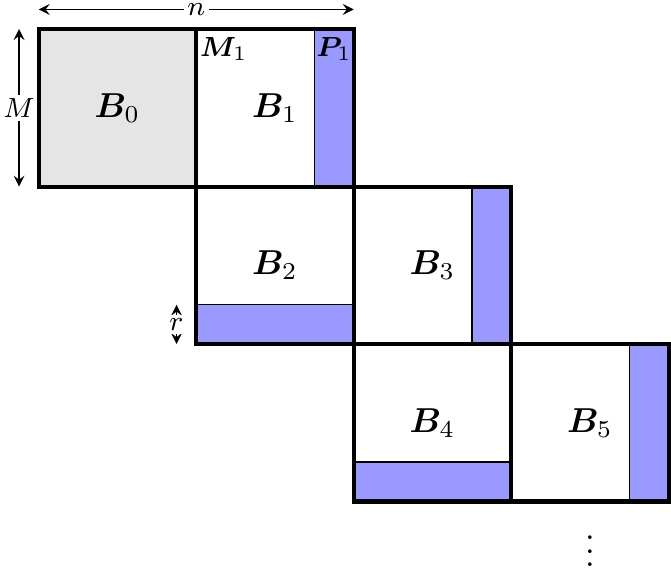}
\caption{Staircase code block structure. Information bits (white) and parity bits (shaded) are shown. Bits in block $\bB^T_0$ are fixed.}
\label{fig:staircase}
\end{figure}

An illustration of a staircase code is shown in Fig. \ref{fig:staircase}. The fundamental building block is a binary, linear, systematic block code $C(n,k)$, referred to as a \emph{component code}, with block-length $n$ (required to be even) and number of information bits $k$. Let $R_c \triangleq k/n$ be the component code rate. For $M \triangleq n/2$, the dimension of each staircase block $\bB_i$ is $M \times M$. For a staircase code to have non-trivial rate (i.e., $R>0$) the component code rate must satisfy $R_c>1/2$.

The first staircase block $\bB_0$ is fixed to all-zero bit-values. Let $r\triangleq n-k$ be the number of parity bits in a component codeword. Let $\bG$ be a $k \times n$ systematic generator matrix for $C$. We denote by $\bG_p$ the $k \times r$ sub-matrix of $\bG$ containing the $r$ columns which correspond to the parity bits in each codeword. For $i\in \{1,2,\dots\}$, given block $i-1$, to encode the $i$th block, first fill an $M \times (M-r)$ matrix $\bM_i$ with information bits. Next, calculate the $M \times r$ matrix $\bP_i$ of parity bits according to
\begin{equation}\label{eqn:sc_encode}
\bP_i = \begin{bmatrix} \bB_{i-1}^T & \bM_i\end{bmatrix}\bG_p
\end{equation}
where $()^T$ denotes matrix transpose. The $i$th block is then given by $\bB_i = \begin{bmatrix}\bM_i & \bP_i\end{bmatrix}$.

The rate of a staircase code is given by
\begin{equation}\label{eqn:rate_sc}
R = 2R_c-1,
\end{equation}
where we assumed that the smallest transmission granularity is a complete block $\bB_i$.

\subsection{Decoding of staircase codes}

Staircase codes are decoded using a \emph{sliding-window decoder}. Consider the blocks in Fig. \ref{fig:staircase} now to be received blocks buffered in a decoding window of length 6, with all except $B_0$ corrupted by a BSC. 

Decoding proceeds in iterations. Let $l\in\{1,2\dots,l_{\textrm{max}}\}$ denote decoding iterations, with the maximum number of iterations denoted by $l_{\textrm{max}}$. During iteration $l$, for each $i\in\{1,2,\dots,5\}$, form the matrix $\begin{bmatrix}\bB_{i-1}^T & \bB_i\end{bmatrix}$ and decode each row of the matrix by a component code decoder, e.g., a syndrome decoder. Once $l=l_{\textrm{max}}$ is reached, the window ``slides'' by shifting out decoded block $\bB_0$ and shifting in a newly received block $\bB_6$. The decoding process continues indefinitely in this manner.

In practice, the component code decoder can be implemented using efficient table-lookup methods for syndrome decoding to achieve very high decoding throughputs \cite[Appendix]{smith:2012}\cite{chien:1969}.

\subsection{Motivation}

Substituting $\bB_{i-1}=\begin{bmatrix}\bM_{i-1} & \bP_{i-1}\end{bmatrix}$ into (\ref{eqn:sc_encode}), we obtain
\[
\bP_i = \begin{bmatrix} \begin{bmatrix}\bM_{i-1} & \bP_{i-1}\end{bmatrix}{}^T & \bM_i\end{bmatrix} \bG_p
\]
which is a linear recursion of the parity-bit matrix $\bP_i$. We refer to this as the \emph{parity-propagation} property of staircase codes. The presence of feedback in the encoding process leads to a number of issues, the most significant of which is the lack of a termination mechanism.

Although staircase codes were designed for continuous transmission applications where termination is not necessary, allowing the encoding process to terminate after a certain number of blocks would extend their domain of application significantly. Furthermore, a terminated staircase code can be decoded by two sliding window decoders working in parallel from both ends of the code. The decoding throughput is doubled at a cost of extra hardware, a favorable trade-off in high-throughput optical-fiber systems.

% other motivating factors...

% Section: Feed-forward staircase code
\section{Feed-forward staircase code}\label{sec:ff_sc}

\begin{figure}
\centering
\includegraphics{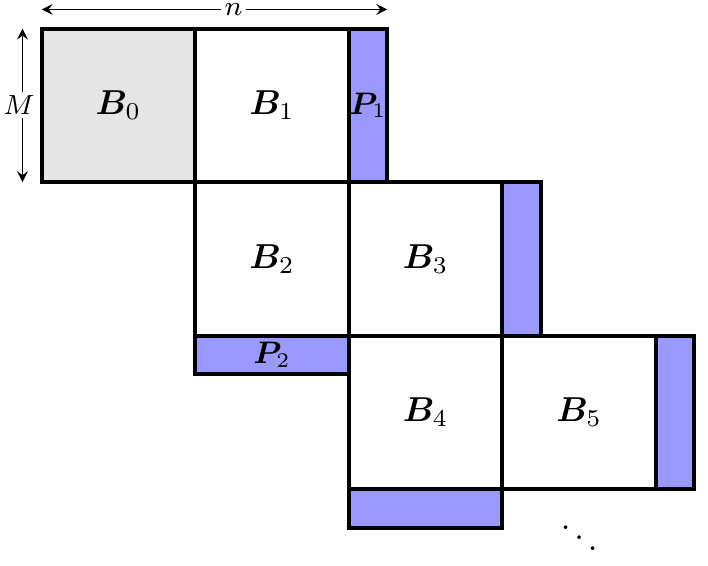}
\caption{Staircase codes without parity re-encoding. Information bits (white) and parity bits (shaded) are shown. Parity-bits are not used for re-encoding.}
\label{fig:ff_staircase_simple}
\end{figure}

The most pragmatic approach to mitigate the effect of parity propagation would be to not re-encode the parity bit block $\bP_i$. Such an approach is shown in Fig.~\ref{fig:ff_staircase_simple}. However, it becomes quickly obvious that this approach suffers from some important problems. Most importantly, if high-rate component codes with error correcting capability $t$ are used, the occurrence of $t+1$ errors in the parity-part of a component code will not be corrected. Hence, if there are $t+1$ errors in the parity part of a vertical codeword, $t+1$ errors in the parity part of a horizontal codeword and an additional error in the intersection of both vertical and horizontal codewords, this additional error will not be corrected and will contribute to the error floor of the code, which will become unacceptably high for most applications. Especially in optical communications, where usually residual bit error rates in the range of $10^{-13}$ to $10^{-15}$ are required, a different approach is necessary.

\begin{figure}
\centering
\includegraphics{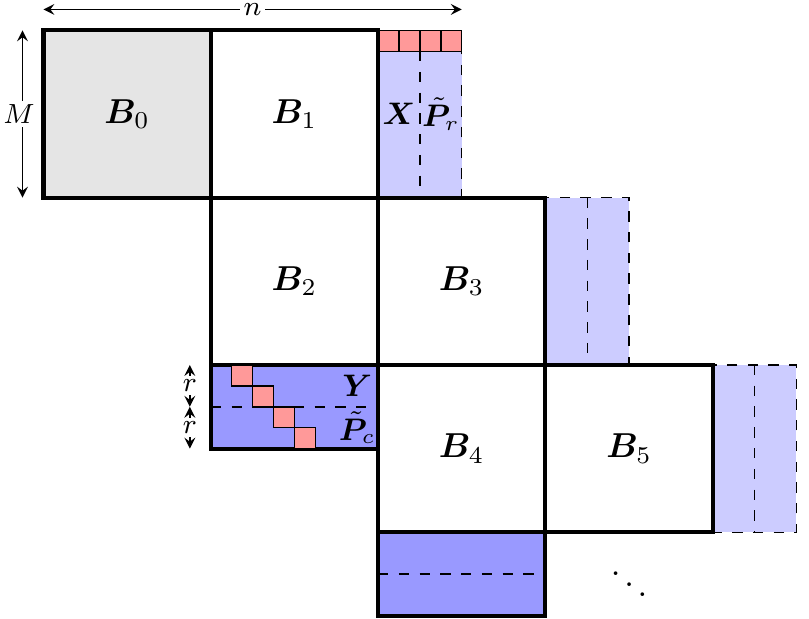}
\caption{Proposed feed-forward staircase code block structure. Information bits ($\bB_i$, white) and column redundancy bits ($\bY$, $\tilde{\bP}_c$, shaded dark) are transmitted. Row redundancy bits ($\bX$, $\tilde{\bP}_r$, shaded light) are punctured. Bits in block $\bB_0$ are fixed. The small squares illustrate permutation selected for low error-floors.}
\label{fig:ff_staircase}
\end{figure}

In order to design a code with acceptable error floors, we adopt the parity self-protection technique proposed in~\cite[App. I.9]{g.975.1:2004} to ensure that errors in the parity part of the code do not cause large residual error floors.
The structure of the proposed Feed-Forward Staircase Code (FF-SC) with parity self-protection is shown in Fig.~\ref{fig:ff_staircase}. The dark shaded blocks at the bottom of even-indexed information blocks are referred to as column \emph{redundancy blocks}. Each column redundancy block consists of a parity block $\tilde{\bP}_{c}$ and a \emph{self-protection} block $\bY$. The lightly shaded blocks to the right of odd-indexed information blocks are referred to as row redundancy blocks, each consisting of a parity block $\tilde{\bP}_{r}$ and a self-protection block~$\bX$, which are both punctured (indicated by the light shading in Fig.~\ref{fig:ff_staircase}). 

As in a staircase code, an FF-SC parity block contains parity bits calculated during component code encoding. The key difference in an FF-SC is that the bits in a self-protection block, which are a sub-set of the information bits of component codes, are additionally constrained. 

Let $\pi_1$ and $\pi_2$ be permutations defined by
\[
\pi_b(\bA) \triangleq \textrm{vec}^{-1}(\bPi_b\textrm{vec}(\bA))
\]
where $b\in\{1,2\}$, $\bA$ is an $M \times r$ matrix, and $\bPi_b$ is an $Mr \times Mr$ permutation matrix. By definition, $\pi_b$ are bijective maps, with the property $\pi_b(\bA+\bB) = \pi_b(\bA) + \pi_b(\bB)$.

We define the \emph{self-protection constraints}
\begin{align}
\bY &= (\pi_1(\bX))^T \label{eqn:self_protection1} \\
\tilde{\bP}_c &= (\pi_2(\tilde{\bP}_r))^T \label{eqn:self_protection2}.
\end{align}
Since $\pi_b$ are bijective, as long as the self-protection constraints are satisfied, we can puncture either the column or row redundancy blocks. For consistency with Fig \ref{fig:ff_staircase}, we puncture the row redundancy blocks in the following.

Due to the constraints imposed on self-protection blocks, $M$ must satisfy $2M + r=k$, hence $M = (k-r)/2$ (assuming $k$ and $r$ have the same parity, which can be achieved with shortening). For computing the rate, we first assume that always an even number of blocks $\bB_i$ are transmitted as smallest granularity. The rate of an FF-SC is then
\begin{equation}\label{eqn:rate_ff}
R_{\textrm{FF}} = 2R_c-1 = R,
\end{equation}
which is identical to the rate of a staircase code. If we want to achieve the finer granularity of conventional staircase codes with single blocks, we define that the parity and self-protection blocks $\bY$ and $\tilde{\bP}_c$ are attached to each block with odd index. In that case, with a total of $\Lambda$ blocks transmitted we have
\begin{equation*}
R_{\textrm{FF}}^\prime = \frac{2k-n}{2k-n+4\lfloor\frac{\Lambda+1}{2}\rfloor\frac{1}{\Lambda}(n-k)}\,,
\end{equation*}
which takes into account the potential transmission of an odd number of blocks.
As $\limsup_{\Lambda\to\infty}\lfloor\frac{\Lambda+1}{2}\rfloor\frac{1}{\Lambda} = \liminf_{\Lambda\to\infty}\lfloor\frac{\Lambda+1}{2}\rfloor\frac{1}{\Lambda} = \frac{1}{2}$, we get
\begin{equation*}
\lim_{\Lambda\to\infty} R_{\textrm{FF}}^\prime = \frac{2k-n}{2k-n+2(n-k)} = R_{\textrm{FF}}\,.
\end{equation*}

\subsection{Encoding}

We slightly generalize the component code definition to allow \emph{different} binary linear block codes to be used as row and column component codes. Given block-length $n$ and number of information bits $k$, let $C_r(n,k)$ be a row component code with $k \times n$ systematic generator matrix $\bG$. Let $C_c(n,k)$ be a column component code with $k \times n$ systematic generator matrix $\bF$. Let $\bG_p$ and $\bF_p$ denote the sub-matrices containing the $r$ columns of $\bG$ and $\bF$ corresponding to parity-bits. 

Due to the self-protection block, the last $r$ bits out of $k$ information bits in a component codeword are constrained. We highlight this fact by partitioning $\bG_p$ and $\bF_p$ according to 
\[
\bG_p = \left[\begin{array}{c} \bG_i \\ \bG_r \end{array}\right]
\;
\bF_p = \left[\begin{array}{c} \bF_i \\ \bF_r \end{array}\right],
\]
where $\bG_i$ and $\bF_i$ are $(k-r) \times r$ matrices and $\bG_r$ and $\bF_r$ are $r \times r$ matrices.

Consider the encoding operation over information blocks $\bB_0$, $\bB_1$, and $\bB_2$ in Fig. \ref{fig:ff_staircase}. Subsequent blocks are encoded in the same manner. By horizontally concatenating $\bB_0$ and $\bB_1$, we obtain 
\[
\bP_r=\begin{bmatrix}\bB_0 & \bB_1\end{bmatrix}\bG_i.
\]
By vertically concatenating $\bB_1$ and $\bB_2$, we obtain
\[
\bP_c=\bF_i^T\begin{bmatrix} \bB_1 \\ \bB_2 \end{bmatrix}.
\]
Note that $\bP_r$ and $\bP_c$ are not the same as $\tilde{\bP}_r$ and $\tilde{\bP}_c$. 

Consider the entries of the $M\times r$ matrix $\bX$ and the $r\times M$ matrix $\bY$ to be variables. According to the structure shown in Fig.~\ref{fig:ff_staircase}, we can write $\tilde{\bP}_r$ and $\tilde{\bP}_c$ as
\[
\tilde{\bP}_r = \bP_r + \bX\bG_r,
\qquad
\tilde{\bP}_c = \bP_c + \bF_r^T\bY.
\]
Imposing self-protection conditions (\ref{eqn:self_protection1}) and (\ref{eqn:self_protection2}), we obtain
\[
\bP_c+(\pi_2(\bP_r))^T = \bF_r^T\bY+(\pi_2(\pi_1^{-1}(\bY^T)\bG_r))^T.
\]

Each of the above terms is an $r\times M$ matrix. Let $\textrm{vec}(\cdot)$ be the column-wise vectorization and let $\bm{y}=\textrm{vec}(\bY)$, $\bm{p}_c=\textrm{vec}(\bP_c)$, and $\bm{p}_r=\textrm{vec}((\pi_2(\bP_r))^T)$. Let $\bPi_T$ be the permutation matrix satisfying $\bY^T=\textrm{vec}^{-1}(\bPi_T\textrm{vec}(\bY))$. Using the fact that for some matrix $\bQ$
\[
\textrm{vec}(\bQ\bY) = (\bI_M \otimes \bQ) \textrm{vec}(\bY),
\]
the above expression can be written as
\begin{multline*}
\bm{p}_c + \bm{p}_r = \\
[\bI_M \otimes \bF_r^T + \bPi_T\bPi_2\bPi_T (\bI_M \otimes \bG_r^T) \bPi_T\bPi_1^{-1}\bPi_T]\bm{y}
\triangleq \bA\bm{y}.
\end{multline*}
If $\bA$ is invertible, then the matrix $\bY$ is given by 
\begin{align}
\bm{y} &= \bA^{-1}(\bm{p}_c + \bm{p}_r) \nonumber \\
&\triangleq \bA^{-1}\bm{c}. \label{eqn:ff_encode}
\end{align}

The invertibility of $\bA$ depends on the choices of $C_r(n,k)$, $C_c(n,k)$, $\bF$, $\bG$, $\bPi_1$, and $\bPi_2$. Using the same row and column component codes, we have found that searching over the space of all $\bPi_1$ and $\bPi_2$ can quickly produce an invertible $\bA$. The search and calculation of $\bA^{-1}$ can be performed offline at design time, since information bits are only involved in the calculation of $\bm{c}$. 

The main complexity of FF-SC encoding is the multiplication in (\ref{eqn:ff_encode}) between an $Mr \times Mr$ matrix and an $Mr \times 1$ vector. The complexity of this operation highly depends on the choice of permutation matrices $\bPi_1$ and $\bPi_2$. For instance, the permutation matrices may be chosen such that the hardware implementation is simplified or such that $\bA^{-1}$ has a special structure easing the multiplication.

\subsection{Decoding}

Decoding of FF-SC is very similar to conventional staircase codes. A sliding window decoder is used starting from block $\bB_0$. When corrections are made in a column redundancy block the corresponding row redundancy block is also modified, and vice versa. Additional logic is required to implement the permutations $\pi_1$, $\pi_2$, and their inverses.

\subsection{Low error-floor permutations}\label{subsec:low_ef}

We describe a choice of permutations $\pi_1$ and $\pi_2$ suitable for applications requiring very low error-floors. The permutations $\pi_1$, $\pi_2$ are defined by the permutation matrices 
\begin{align*}
\bPi_1 &= \bm{\mathcal{B}}(\bE_{M}^{M-1},\bE_{M}^{M-2},\dots,\bE_{M}^{M-r}) \\
\bPi_2 &= \bm{\mathcal{B}}(\bE_{M}^{M-r-1},\bE_{M}^{M-r-2},\dots,\bE_{M}^{M-2r}),
\end{align*}
together with column-wise vectorization $\textrm{vec}(\cdot)$ and its inverse $\textrm{vec}^{-1}(\cdot)$. 

These permutations cyclically shift each column of $\bX$ and $\tilde{\bP}_r$ by a number of bits related to their column index, an example of which is shown in Fig. \ref{fig:ff_staircase}. They can be implemented efficiently in hardware using barrel shifters. Discussions of the estimated and simulated error-floor performance under these permutations are given in Sec.~\ref{subsec:error_floor}.

\section{Partial Feed-forward staircase code}\label{sec:pff_sc}

\begin{figure*}
\centering
\includegraphics{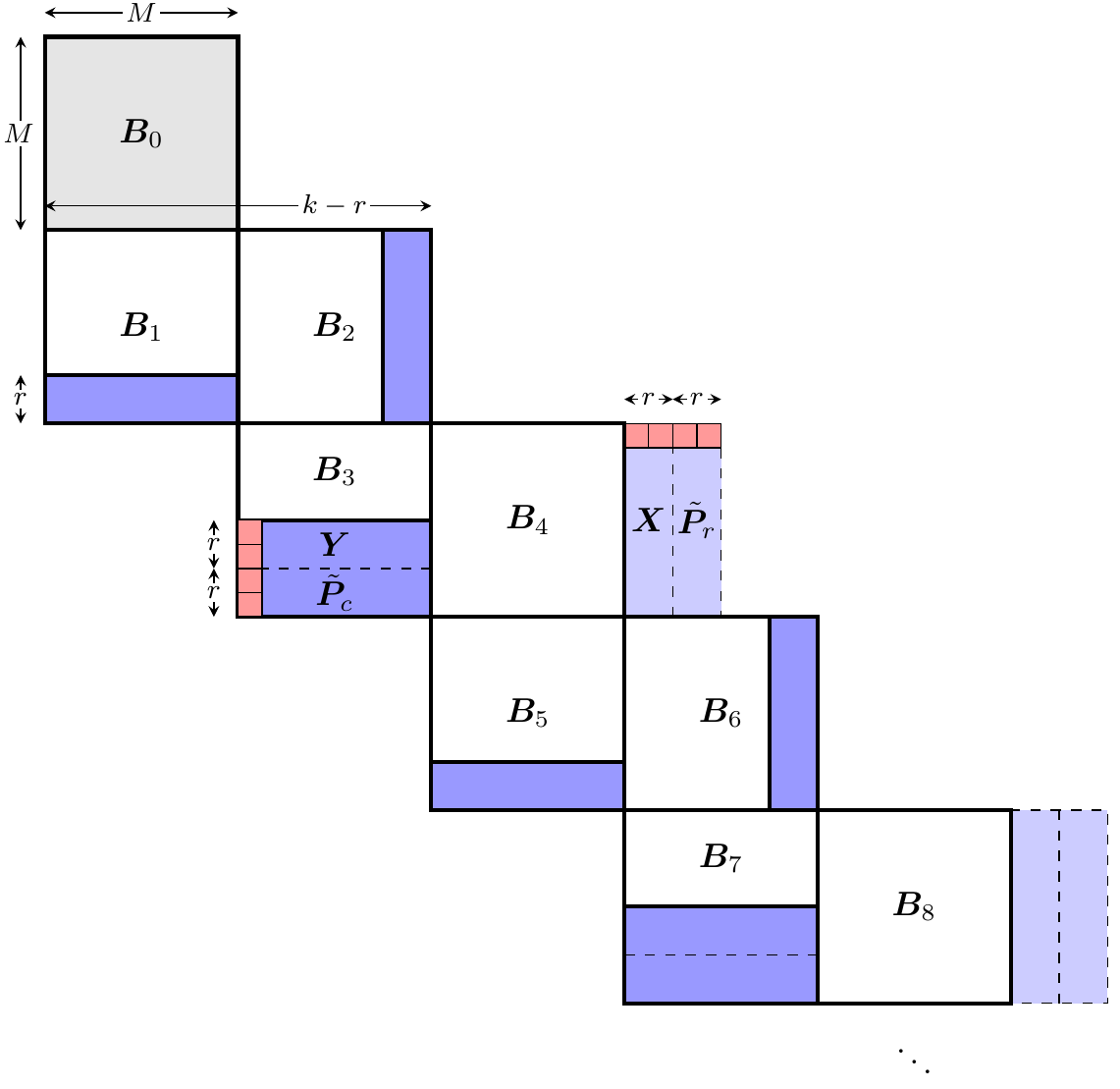}
\caption{Partial feed-forward staircase code block structure. Information bits ($\bB_i$, white), parity bits (shaded dark), and column-redundancy bits ($\bY$, $\tilde{\bP}_c$, shaded dark) are transmitted. Row redundancy bits ($\bX$, $\tilde{\bP}_r$, shaded light) are punctured. Bits in block $\bB_0$ are fixed. The small squares illustrate the trivial permutations.}
\label{fig:pff_staircase}
\end{figure*}

Although self-protection allows us to considerably reduce the error floor of feed-forward staircase codes, the error floor (see Sec.~\ref{sec:simulations}) may still be unacceptably high for some applications requiring very low residual BERs, e.g., optical core networks. We therefore slightly relax the parity-propagation constraint by allowing the parity bits to propagate over some blocks and introduce \emph{Partial Feed-Forward Staircase Codes} (PFF-SCs). 

Let $L \in \{1,2,\dots\}$ be the \emph{propagation length} of a PFF-SC, defined as the maximum number of consecutive blocks over which parity-propagation can occur. The PFF-SC then uses a hybrid structure, with $L-1$ blocks being standard staircase code blocks followed by one block with parity bits that are not re-encoded but where self-protection is used to mitigate the detrimental effect of harmful error patterns. The self-protection scheme also results in one block containing only information bits. Figure~\ref{fig:pff_staircase} illustrates the structure of a PFF-SC with $L=3$. In this example, $2$ out of every $4$ blocks are standard staircase code blocks and $1$ out of every $4$ blocks contains only information bits. Self-protection is used to stop parity-propagation after $L=3$ blocks.

Another major difference in PFF-SCs is the position of the self-protection redundancy blocks, which are part of the conventional staircase structure. This modification allows the permutations $\pi_1$, $\pi_2$ to be trivial and drastically reduces the error-floor as compared to FF-SC (see Sec. \ref{sec:simulations}). Another difference is that the number of information bits per block $\bB_i$ is not constant. As in FF-SC, we set $M=(k-r)/2$ to account for the self-protection and all blocks contain $M^2$ code bits. The component codes are shortened respectively. In order to accommodate the position of self-protection redundancy blocks $\bY$, the component codes involved in self-protection (e.g., codes over blocks $\bB_2$ and $\bB_3$ as well as $\bB_6$ and $\bB_7$ in Fig.~\ref{fig:pff_staircase}) must be shortened by an extra $2r$ bits relative to the other component codes. 

\subsection{Rate of PFF-SCs}
In order to compute the rate of PFF-SCs, we count the number of information bits per block. The first $L-1$ blocks $\bB_{1+(L+1)i},\ldots \bB_{L-1+(L+1)i}$, $i \in \{0,1,\ldots\}$ out of $L+1$ blocks (e.g, $\bB_1$ and $\bB_2$ in Fig.~\ref{fig:pff_staircase}) are standard staircase code blocks of size $M\times M$ with $M(M-r) = \frac{1}{4}\left(k^2+3r^2-4kr\right)$ information bits. The block $\bB_{L+(L+1)i}$, $i \in \{0,1,\ldots\}$ contains exactly $M(M-2r) = \frac{1}{4}\left(k^2+5r^2-6kr\right)$ information bits and finally, the block $\bB_{(L+1)(i+1)}$, $i \in \{0,1,\ldots\}$ contains exactly $M^2 = \frac{1}{4}(k-r)^2$ information bits. For computing the rate, we must fix again the granularity of transmission. If we assume that always $L+1$ blocks $\bB_1,\ldots,\bB_{(L+1)i}$, $i\in\mathbb{N}$ are transmitted, then the rate can be computed as
\begin{align}
R_{\textrm{PFF}} &= \frac{(L-1)M(M-r)+M(M-2r)+M^2}{(L+1)M^2}\nonumber\\
&= 1-\frac{r}{M} = 1+\frac{2R_c-2}{2R_c-1},\label{eqn:rate_pff}
\end{align}
which is independent of $L$. As $R_{\textrm{PFF}}-R=\frac{(2R_c-2)^2}{1-2R_c}$, we can conclude that $R_{\textrm{PFF}} < R$ as $R_c > \frac{1}{2}$ has to hold (see Sec.~\ref{subsec:staircase_enc}). However, at high rates the differences are small. For example, $R_{\textrm{PFF}}$ is within $5\%$ of $R$ for $R_c \geq 10/11$ and within $25\%$ for $R_c\geq 5/6$. Note that a PFF-SC of non-trivial rate requires $R_c>3/4$. 

This result may seem counter-intuitive at first glance, since it appears that we should recover the original staircase code rate $R$ for $L\to\infty$. However, contrary to the original staircase code construction (see Sec.~\ref{sec:background}), in the proposed construction the component codes of the staircase-like blocks are shortened by $2r$, which leads to the observed rate difference. We could relax the granularity constraint of $L+1$ blocks and find an expression for $R_{\text{PFF}}^\prime(\Lambda,L)$. As this expression is cumbersome and does not lead to any new insights, we omit it here. For practical purposes, it is customary to restrict ourselves to the granularity of $L+1$ blocks, allowing for easy termination and avoiding possibly higher error rates at the code boundaries.

\subsection{Description of the Encoder}
\begin{figure}
\centering
\includegraphics{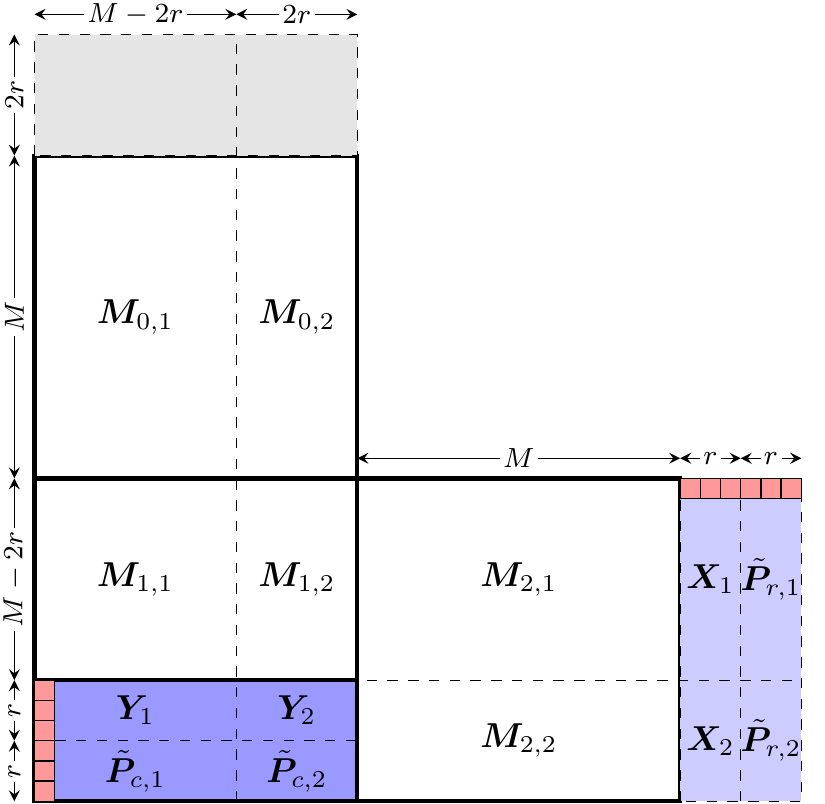}
\caption{Sub-block divisions for PFF-SC encoding. The $2r \times M$ sub-block at the top (shaded light) is shortened.}
\label{fig:pff_encode}
\end{figure}

In this subsection, we describe the encoder of PFF-SC. We focus only on the self-protection blocks, since $L-1$ out of $L+1$ consecutive blocks are encoded in the same way as the original staircase code. Our explanations will focus on Fig.~\ref{fig:pff_encode}, which highlights blocks $\bB_2$, $\bB_3$, $\bB_4$, $\bY$, $\tilde{\bP}_c$, $\bX$, and $\tilde{\bP}_r$ of Fig.~\ref{fig:pff_staircase} for $L=3$.

Figure \ref{fig:pff_encode} further sub-divides each block into sub-blocks. The encoding process consists of two stages. Stage 1 calculates $\bY_1$. Stage 2 calculates $\bY_2$ based on $\bY_1$. In terms of implementation complexity, stage 1 is equivalent to component code encoding while stage 2 is a general matrix multiplication. Fortunately, for high code rates where $M \gg r$, the encoding complexity is dominated by stage 1.

\subsubsection{Calculating $\bY_1$}

We inherit the definitions of matrices $\bG_p$, $\bF_p$, $\bG_i$, $\bF_i$, and $\bG_r$, $\bF_r$ from Sec.~\ref{sec:ff_sc}. By horizontally concatenating $\bM_{1,1}$, $\bM_{1,2}$, and $\bM_{2,1}$, we obtain
\begin{equation}\label{eqn:pr1}
\bP_{r,1}=\begin{bmatrix}\bM_{1,1} & \bM_{1,2} & \bM_{2,1}\end{bmatrix}\bG_i.
\end{equation}
By vertically concatenating blocks $\bm{0}_{2r \times M-2r}$, $\bM_{0,1}$ and $\bM_{1,1}$, where $\bm{0}_{2r \times M-2r}$ accounts for the extra shortening of the column component codes, we obtain
\[
\bP_{c,1}=\bF_i^T\begin{bmatrix} \bm{0}_{2r \times M-2r} \\ \bM_{0,1} \\ \bM_{1,1}\end{bmatrix}.
\]
We write $\tilde{\bP}_{c,1}$ and $\tilde{\bP}_{r,1}$ as
\[
\tilde{\bP}_{c,1} = \bP_{c,1} + \bF_r^T\bY_1,
\qquad
\tilde{\bP}_{r,1} = \bP_{r,1} + \bX_1\bG_r.
\]
Imposing self-protection constraints
\[
\bY_1 = \bX_1^T,
\qquad
\tilde{\bP}_{c,1} = \tilde{\bP}_{r,1}^T
\]
under trivial permutations and solving for $\bY_1$ gives
\begin{align}
\bY_1 &= \left(\bG_r^T + \bF_r^T\right)^{-1}\left(\bP_{c,1} + \bP_{r,1}^T\right) \nonumber \\
%&\triangleq \bA^{-1}\bC. \label{eqn:a1}
&\triangleq \bA^{-1}\left(\bP_{c,1} + \bP_{r,1}^T\right). \label{eqn:a1}
\end{align}

Since $\bA=\bm{0}_{r \times r}$ if $\bG_r=\bF_r$, a necessary condition for $\bA$ to be invertible is $\bG_r \neq \bF_r$. Here we satisfy this condition by using different binary cyclic codes as row and column component codes. However, instead of using different component codes with different error correction capabilities and potentially requiring distinct decoder hardware implementations, we propose the following construction:
Let $g(x)$ and $f(x)$ be generator polynomials for $C_r(n,k)$ and $C_c(n,k)$. We require $g(x)$ and $f(x)$ to satisfy the condition
\begin{equation}\label{eqn:reciprocal}
f(x) = x^{\textrm{deg}(g(x))}g(x^{-1})
\end{equation}
where $\textrm{deg}(p(x))$ is the degree of the polynomial $p(x)$. The component codes then have the property that the ``mirror-image'' of a codeword $(c_0,c_1,\dots,c_{n-1}) \in C_r(n,k)$, i.e., $(c_{n-1},c_{n-2},\dots,c_0)$, is a codeword of $C_c(n,k)$, and vice versa~\cite[Ch. 7]{MacWilliamsSloane}.  Hence, the same decoder hardware can be reused to decode both component codes, with some simple bit-reversal logic.

Using different binary cyclic component codes with generator polynomials satisfying (\ref{eqn:reciprocal}) gives an invertible $\bA$ as $\bG_r \neq \bF_r$. By calculating $\bA^{-1}$ offline at design time, the complexity of finding $\bY_1$ and $\tilde{\bP}_{c,1}$ at encoding time is equivalent to a multiplication between an $r \times r$ matrix and an $r \times M-2r$ matrix.

\subsubsection{Calculating $\bY_2$}

In stage 2, the blocks $\bY_1$ and $\tilde{\bP}_{c,1}$ are considered known. By vertically concatenating blocks $\bm{0}_{2r\times2r}$, $\bM_{0,2}$ and $\bM_{1,2}$ we obtain
\[
\bP_{c,2} = (\bF_i)^T\begin{bmatrix}\bm{0}_{2r\times2r} \\ \bM_{0,2} \\ \bM_{1,2} \end{bmatrix}
\]
hence  
\begin{equation}\label{eqn:pc2}
\tilde{\bP}_{c,2} = \bP_{c,2} + \bF_r^T\bY_2.
\end{equation}
We partition the matrix $\bG_i$ into $3$ sub-matrices with
\[
\bG_i = \begin{bmatrix} \bG_A \\ \bG_B \\ \bG_C \end{bmatrix}
\]
where $\dim\bG_A=(M-2r) \times r$, $\dim\bG_B=2r \times r$, and $\dim\bG_C=M \times r$. We can now write
\[
\tilde{\bP}_{r,2} = \begin{bmatrix} \bY_1 \\ \tilde{\bP}_{c,1} \end{bmatrix} \bG_A +  \begin{bmatrix} \bY_2 \\ \tilde{\bP}_{c,2} \end{bmatrix}\bG_B + \bM_{2,2}\bG_C + \bX_2\bG_r.
\]
Using (\ref{eqn:pc2}) and the self-protection constraint $\bY_2^T = \bX_2$, we have
\[
\tilde{\bP}_{r,2} = \left[ \begin{array}{c} \bY_1 \\ \tilde{\bP}_{c,1} \end{array} \begin{array}{c} \bm{0}_{r\times 2r} \\ \bP_{c,2} \end{array}\  \bM_{2,2}\right]\bG_i + \begin{bmatrix} \bI_r \\ \bF_r^T \end{bmatrix}\bY_2 \bG_B + \bY_2^T\bG_r.
\]
Imposing the self-protection constraint $\tilde{\bP}_{r,2}=(\tilde{\bP}_{c,2})^T$ and simplification yields
\begin{equation}\label{eqn:y2}
\bY_2^T\bA^T + \begin{bmatrix} \bI_r \\ \bF_r^T \end{bmatrix} \bY_2\bG_B = \bC
\end{equation}
where $\bA$ was defined implicitly in (\ref{eqn:a1}) and with
\[
\bC \triangleq \left[ \begin{array}{c} \bY_1 \\ \tilde{\bP}_{c,1} \end{array} \begin{array}{c} \bm{0}_{r\times 2r} \\ \bP_{c,2} \end{array}\ \bM_{2,2}\right]\bG_i + \begin{bmatrix}\bm{0}_{2r\times 2r} & \bM_{0,2}^T & \bM_{1,2}^T\end{bmatrix} \bF_i.
\]
Note that all terms in (\ref{eqn:y2}) are $2r \times r$ matrices. 

Let $\textrm{vec}(\cdot)$ now denote the \emph{row-wise} vectorization given by the mapping $v(i,j) = in+j$. Let $\bm{y}=\textrm{vec}(\bY_2)$ and $\bm{c}=\textrm{vec}(\bC)$. 
Let $\bm{\mathcal{S}}(\bA)$ be the $r \times 2r^2$ matrix where for $i\in[0,r-1]$ and $j=2ri$, the $j$th column of $\bm{\mathcal{S}}(\bA)$ is the $i$th column of $\bA$, with zeros elsewhere. We can then equivalently write (\ref{eqn:y2}) as 
\[
\bB\bm{y} = \bm{c} 
\]
where $\bB$ is the $2r^2 \times 2r^2$ matrix given by
\[
\bB \triangleq \begin{bmatrix} 
\bm{\mathcal{S}}(\bA) \\
\bm{\mathcal{S}}(\bA)\bE_{2r^2} \\
\vdots \\
\bm{\mathcal{S}}(\bA)\bE_{2r^2}^{2r-1}
\end{bmatrix}
+
\begin{bmatrix}
\bI_r \otimes \bG_B^T \\
\bF_r^T \otimes \bG_B^T
\end{bmatrix}.
\]

\subsubsection{Finding an invertible $\bB$}

Since $\bG_r$ and $\bF_r$ were fixed in stage 1 in order to obtain an invertible $\bA$, if $\bB$ is singular, the only way to obtain an invertible $\bB$ is to manipulate $\bG_B$ using elementary row operations. Here we focus on row permutations of $\bG_B$ only, since they do not affect the error floor. 

Let $\bPi$ be a $2r \times 2r$ permutation matrix. Denote the permuted $\bG_B$ by $\tilde{\bG}_B \triangleq \bPi \bG_B$. A computer search can be used to find an appropriate $\Pi$ that results in an invertible $\bB$. 

Given $\bPi$, the expressions for $\tilde{\bP}_{r,2}$ and $\bB$ are modified by replacing $\bG_B$ with $\tilde{\bG}_B$. Note that $\bPi$ also affects stage 1 calculations, where (\ref{eqn:pr1}) has to be modified to
\[
\bP_{r,1}=\begin{bmatrix}\bM_{1,1} & \bM_{1,2}\bPi & \bM_{2,1}\end{bmatrix}\bG_i.
\]
For an invertible $\bB$, the matrix $\bY_2$ is given by 
\[
\bm{y}=\bB^{-1}\bm{c}.
\]

The complexity of calculating $\bY_2$ is dominated by the multiplication with a $2r^2\times 2r^2$ matrix. Since only $1$ out of every $L+1$ blocks requires self-protection calculations, the average complexity of PFF-SC approaches conventional staircase codes with increasing $L$.

\section{Error-floor Analysis}\label{subsec:error_floor}

Error-floor analysis of staircase codes and its variants proposed in this paper depends on enumerating the number of \emph{stall patterns}, i.e., patterns of errors that the decoder cannot remove~\cite{justesen:2011,smith:2012}. To obtain a simple estimate of the error-floor, we only enumerate the smallest stall patterns resulting from channel errors, referred to as \emph{minimal} stall patterns.

We consider an erroneously decoded bit to be a bit error only if it is an information bit. A decoded block is considered to be a block error if it contains at least one bit error. The block (BKER) and bit (BER) error-rates are defined according to these definitions.

\begin{figure}
\centering
\includegraphics{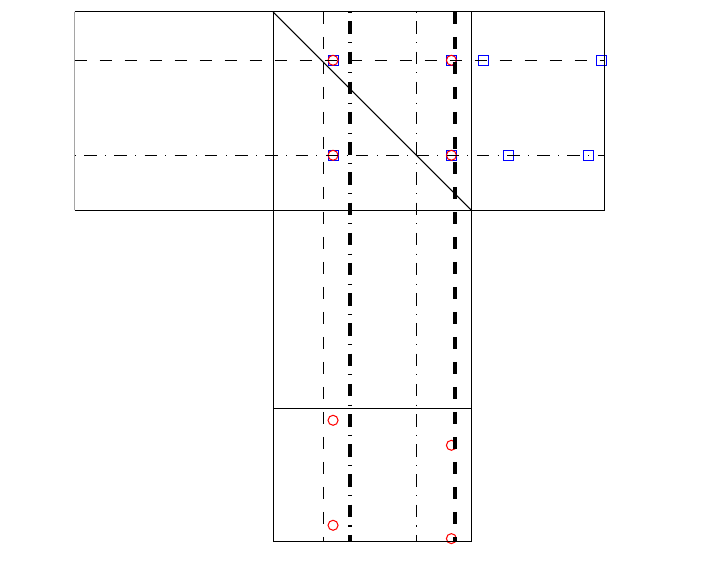}
\caption{Minimal stall pattern used to estimate FF-SC error floors for component codes with $t=3$. Blue (\textcolor{blue}{$\Box$}) markers are bit errors in row component codes. Red (\textcolor{red}{$\circ$}) markers are bit errors in column component codes. Dashed and dash-dotted lines are referred to in the derivation of error-floor estimates in Sec. \ref{subsec:error_floor}. Note that only $4$ out of the $8$ bit errors in redundancy blocks are received from the channel, the other ones are interleaved versions thereof.}
\label{fig:ff_stall_pattern}
\end{figure}

We estimate the block and bit error-floors of FF-SC based on low-error-floor permutations of Sec. \ref{subsec:low_ef} assuming transmission over a BSC with error probability $p$. An example of a minimal stall pattern for component codes with $t=3$ is shown in Fig. \ref{fig:ff_stall_pattern}, consisting of $4$ information-bit errors and $4$ redundancy-bit errors from the channel.

To construct such a stall pattern, first choose any $2$ out of $M$ rows in the information block, such as the rows marked by the horizontal dashed and dash-dotted lines in Fig. \ref{fig:ff_stall_pattern}. Denote the chosen rows by $r_1$ and $r_2$. Under the transposes in (\ref{eqn:self_protection1}) and (\ref{eqn:self_protection2}), the chosen rows are mapped to columns marked by the \emph{thin} vertical dashed and dash-dotted lines, reflected about the diagonal of the information block. 

Under the proposed low-error-floor permutations, bit errors in the row redundancy block are cyclically shifted by no more than $2r-1$ columns, modulo $M$, in the column redundancy block. In Fig.~\ref{fig:ff_stall_pattern}, the range of cyclic shifts is bounded by the thin and corresponding \emph{thick} vertical lines. For example, bit errors in the row redundancy block of $r_1$ may be shifted to columns within the thin and thick dashed vertical lines. For $r_2$, bit errors in the row redundancy block may be shifted to columns within the thin and thick dash-dotted vertical lines, wrapping around the right boundary of the column redundancy block.  Given $r_i$, we define its \emph{valid column set} by
\[
S(r_i) \triangleq \{ r_i + j \textrm{ mod } M \textrm{ for all } j\in[0,2r-1] \}.
\]

It is simple to verify the following \emph{spreading property} of the low-error-floor permutations: if $2r<M$, i.e., $R>1/2$ or $OH<100\%$ (where $OH$ denotes the overhead of the code, defined as $OH \triangleq (1/R - 1) \times 100\%$), then row redundancy block bit-errors belonging to the same row \emph{cannot} belong to the same column in the column redundancy block. Consequently, columns in the stall pattern can only be chosen from the \emph{intersection} of valid column sets. The number of choices of such columns is
\begin{equation}
|S(r_1) \cap S(r_2)| \leq 2r.\label{eq:errorfloor_ubs}
\end{equation}
In Fig. \ref{fig:ff_stall_pattern}, the intersection consists of columns bounded between the thin dashed and thick dash-dotted vertical lines and columns bounded between the thin dash-dotted and thick dashed vertical lines. The resulting error-floor estimates based on the simple upper-bound~\eqref{eq:errorfloor_ubs} are given by
\begin{align*}
\textrm{BKER}_{\textrm{FF}} &\approx \binom{M}{2}\binom{2r}{2}p^8 \\
\textrm{BER}_{\textrm{FF}} &\approx  \textrm{BKER}_{\textrm{FF}}\frac{4}{M^2}.
\end{align*}
where $p$ denotes the error probability of the BSC.

For arbitrary $t$, let $t_i = \lfloor(t+1)/2\rfloor$ and $t_r=t+1-t_i$. For odd $t$, $t_i=t_r$ and the above argument for $t=3$ applies directly. Observe that $t_i$ (resp. $t_r$) is then the number of information (resp. redundancy) block bit-errors in each row of a minimal stall pattern. The error-floor estimates for odd $t$ are given by
\begin{align}
\textrm{BKER}_{\textrm{FF}} &\approx \binom{M}{t_r}\binom{2r}{t_r}p^{t_r(t+1)} \label{eqn:ff_sc_bker} \\
\textrm{BER}_{\textrm{FF}} &\approx  \textrm{BKER}_{\textrm{FF}}\frac{t_it_r}{M^2} \label{eqn:ff_sc_ber}.
\end{align}

For even $t$, we first choose $t_i$ rows out of $M$ in the information block. Each erroneous row is assumed to contain $t_i$ bit errors in the information block and $t_r$ bit errors in the row redundancy block. Under the spreading property, bit errors in the row redundancy block are spread to at least $t_r$ distinct columns in the column redundancy block. In the minimal stall pattern, there are \emph{exactly} $t_r$ erroneous columns in the column redundancy block, each containing $t_i$ bit-errors (since the total number of bit errors in the row redundancy block is $t_it_r$). Consequently, there must be $t_r$ erroneous columns in the information block, each containing at least $t+1-t_i=t_r$ bit-errors. We add one additional erroneous row, with $t_i$ bit errors in the information block and $t_r$ bit errors in the row redundancy block, to complete the minimal stall pattern.

The resulting minimal stall pattern contains $t_it_r$ bit errors in the information block and $t_r^2$ bit-errors in the row (or column) redundancy block for a total of $t_r(t_i+t_r)=t_r(t+1)$ bit errors. Applying the intersection of valid column sets argument for the number of choices of columns in the stall pattern, we conclude that the error-floor estimates for even $t$ are also given by (\ref{eqn:ff_sc_bker}) and (\ref{eqn:ff_sc_ber}).

We estimate the block and bit error-floors of PFF-SC based on the minimal stall pattern of weight $16$, with all $16$ bits being information bits. This is the same minimal stall pattern as in the original staircase codes \cite{smith:2012}, obtained by choosing $t+1$ rows out of $M$ followed by $k$ columns out of $M$ in one block and $t+1-k$ columns out of $M$ in the adjacent block, for all $k\in[0,3]$. The error-floor estimates for general $t$ are given by
\begin{align*}
\textrm{BKER}_{\textrm{PFF}} &\approx \binom{M}{t+1} \sum_{k=0}^t\binom{M}{k}\binom{M}{t+1-k}p^{(t+1)^2} \\
\textrm{BER}_{\textrm{PFF}} &\approx \textrm{BKER}_{\textrm{PFF}}\frac{(t+1)^2}{M^2}.
\end{align*}

\section{Simulation Example}\label{sec:simulations}

In this section, we consider FF-SC and PFF-SC based on shortened primitive BCH component codes. Let $m>0$ be the \emph{degree of the extension field} and $t>0$ be the \emph{unique decoding radius} of a primitive BCH code. Let $s\geq 0$ be the number of bits to shorten each BCH code in order to obtain a component code with block-length $n$ and number of information bits $k$. Given $n$ and $k$, the values of $m$, $t$, and $s$ are determined by the constraints 
\[
n=2^m-1-s,\quad k=n-mt.
\]
For fixed $t$, we always choose the smallest $m$ that satisfies these constraints.

Given $t$ and the primitive element $\alpha \in \text{GF}(2^m)$, the row generator polynomial is given by $g(x) = \prod_{i\in[1,2t]}M_{\alpha^i}(x)$ where $M_{\alpha^i}(x)$ is the minimal polynomial of $\alpha^i$. The column generator polynomial, which we choose to be the reciprocal polynomial of $g(x)$, is given by $f(x) = \prod_{i\in[1,2t]}M_{\alpha^{-i}}(x)$ where
\[
M_{\alpha^{-i}}(x) \triangleq x^{\textrm{deg}\left(M_{\alpha^i}(x)\right)}M_{\alpha^i}(x^{-1}).
\]

We constructed FF-SC and PFF-SC of rates $R \in \{3/4,4/5,5/6,13/14\}$. The code parameters are shown in Tables \ref{tbl:ff} and \ref{tbl:pff}. We chose $t=3$ so that error-floors can be studied in the simulation. Furthermore, the selection of $t=3$ yields a very efficient decoder based on lookup tables~\cite{smith:2012}.

\begin{table}
\caption{Feed-forward staircase code parameters}
\label{tbl:ff}
\centering
\begin{tabular}{c|c|c|c|c|c}
$R$ & $OH (\%)$ & $m$ & $t$ & $s$ & $M$ \\ \hline 
3/4 & 33.3 & 8 & 3 & 63 & 72 \\
4/5 & 25.0 & 8 & 3 & 15 & 96 \\
5/6 & 20.0 & 9 & 3 & 187 & 135 \\
13/14 & 7.69 & 10 & 3 & 183 & 390 \\
\end{tabular}
\end{table}

\begin{table}
\caption{Partial feed-forward staircase code parameters}
\label{tbl:pff}
\centering
\begin{tabular}{c|c|c|c|c|c|c|c|c}
$R$ & $OH$ (\%) & $m$ & $t$ & $s$ & $M$ & $p_{15}$ & $\Delta$ & $\Delta_{\textrm{ref}}$ \\ \hline 
3/4 & 33.3 & 8 & 3 & 15 & 96 & $1.82 \cdot 10^{-2}$ & 1.64 & 1.38 \\
4/5 & 25.0 & 9 & 3 & 187 & 135 & $1.56 \cdot 10^{-2}$ & 1.25 & 1.06 \\
5/6 & 20.0 & 9 & 3 & 133 & 162 & $1.30 \cdot 10^{-2}$ & 1.07 & 0.92 \\
13/14 & 7.69 & 10 & 3 & 123 & 420 & $4.80 \cdot 10^{-3}$ & 0.73 & 0.48 \\
\end{tabular}
\end{table}

\begin{figure}
\centering
\includegraphics[width=\columnwidth]{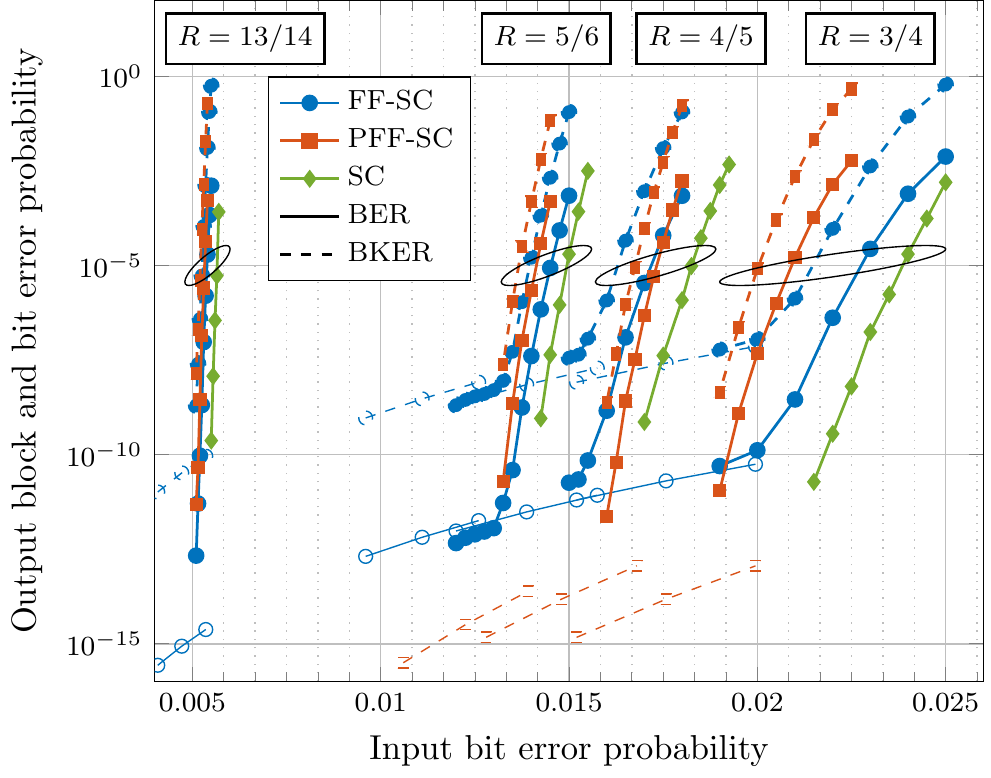}
\caption{Block (dashed lines) and bit-error probabilities (solid lines) of feed-forward (\textbullet) and partial feed-forward ($\blacksquare$) staircase codes with parameters in Tables \ref{tbl:ff} and \ref{tbl:pff}. For reference, conventional staircase codes ($\blacklozenge$) are also shown. Block and bit error-floor estimates are also shown (thin lines, open markers).}
\label{fig:sim}
\end{figure}

Software simulated block and bit error-probabilities of transmission over a BSC using the codes of Tables \ref{tbl:ff} and \ref{tbl:pff} are shown in Fig.~\ref{fig:sim}, along with their error-floor estimates (shown as thin lines with open markers). All FF-SCs were implemented using the low-error-floor permutations of Sec. \ref{subsec:low_ef}. All PFF-SCs were implemented with $L=1$.

Both proposed classes of codes show similar performance in the waterfall region. PFF-SCs have a slight performance loss at lower rates due to their rate loss, which requires a larger $M$ compared to an FF-SC of the same rate.

In the error-floor region, even with low-error-floor permutations, FF-SCs have observable error-floors. On the other hand, PFF-SCs, due to their similarity to the structure of the original staircase codes, do not exhibit any bit error-floor above a BER of $10^{-15}$. In fact, the estimates of the bit error-floor are orders of magnitude below $10^{-15}$. For comparison, we also give the bit error rates of the original staircase codes ($\blacklozenge$) constructed using the same component codes. We can see that the original staircase code slightly outperforms the FF-SC and PFF-SC, especially for low rates, however, at high rates, the difference becomes negligible. This difference is most likely due to the stronger coupling between blocks in the original staircase code construction.

Let $h(x)$ be the binary entropy function and $\textrm{erfc}^{-1}(x)$ be the inverse complementary error function. Given a code of rate $R$ which achieves an output BER of $10^{-15}$ at an input BER of $p_{15}$, we define the NCG gap to capacity (in dB) by
\[
\Delta \triangleq 20\log_{10}\textrm{erfc}^{-1}(2h^{-1}(1-R)) - 20\log_{10}\textrm{erfc}^{-1}(2p_{15})
\]
where $h^{-1}(x)$ is the unique $0 \leq p < 1/2$ such that $h(p)=x$.

We extrapolate the BER curves of PFF-SC down to $10^{-15}$ in order to estimate $p_{15}$. The values of $p_{15}$ with the corresponding $\Delta$ are given in Table \ref{tbl:pff}. For comparison, we also included the $\Delta_{\textrm{ref}}$ of staircase codes of the same rates from~\cite{zhang:2014}, which were found by exhaustively searching over a wide range of parameters $m$ and $t$ and are considered to be the best staircase codes based on the construction given in Sec. \ref{sec:background} and \cite{smith:2012}. The referenced codes were based on BCH component codes with $t\in\{4,5\}$. Nevertheless, the difference in NCG between PFF-SCs with $t=3$ and the reference codes are less than $0.26$\,dB. Error-floors of PFF-SCs and the reference codes are identical and well below $10^{-15}$.

\section{Conclusions}\label{sec:conclusions}

In this paper, we proposed two modifications to staircase codes which allow for convenient termination. In feed-forward staircase codes, a self-protection technique is used to completely eliminate parity-propagation. In partial feed-forward staircase codes, a propagation-length parameter is used to control the extent of parity-propagation. 

Analysis and simulation results show that these codes have similar performance as the original staircase codes. FF-SC have slightly better waterfall performance than PFF-SC, while PFF-SC have much lower error-floors. Hence, FF-SC and PFF-SC are good staircase code solutions for applications where parity-propagation is undesirable or termination is necessary.

\end{document}